\documentclass[aps,prb,twocolumn,showpacs,floatfix]{revtex4}
\usepackage{amsmath,amssymb,amsfonts}
\usepackage{graphics}

\begin{document}

\title{Structural study of an amorphous NiZr$_2$ alloy by 
anomalous wide angle X-ray scattering and Reverse Monte Carlo simulations}

\author{J. C. de Lima}
\email{fsc1jcd@fisica.ufsc.br}
\altaffiliation[Permanent address: ]{Departamento de F\'{\i}sica, Universidade Federal de Santa Catarina, 88040-900 
Florian\'opolis, SC, Brazil}
\affiliation{LURE B\^at. 209D, Facult\'e des Sciences, Orsay 91405, France}

\author{K. D. Machado}
\email{kleber@fisica.ufsc.br}
\affiliation{Departamento de F\'{\i}sica, Universidade Federal de Santa Catarina, 88040-900 
Florian\'opolis, SC, Brazil}

\author{T. A. Grandi}
\affiliation{Departamento de F\'{\i}sica, Universidade Federal de Santa Catarina, 88040-900 
Florian\'opolis, SC, Brazil}
\author{C. E. M. de Campos}
\affiliation{Departamento de F\'{\i}sica, Universidade Federal de Santa Catarina, 88040-900 
Florian\'opolis, SC, Brazil}

\author{D. Raoux}
\affiliation{LURE B\^at. 209D, Facult\'e des Sciences, Orsay 91405, France}
\author{J. M. Tonnerre}
\affiliation{LURE B\^at. 209D, Facult\'e des Sciences, Orsay 91405, France}
\author{D. Udlon}
\affiliation{LURE B\^at. 209D, Facult\'e des Sciences, Orsay 91405, France}

\author{T. I. Morrison}
\affiliation{CSRRI, 3101 South Dearborn Street Room 166 Life Sciences, Chicago, IL 60616}

\date{\today}

\begin{abstract}

The local atomic structure of an amorphous NiZr$_2$ alloy was investigated using the anomalous 
wide-angle x-ray scattering (AWAXS), differential anomalous scattering (DAS) and reverse Monte 
Carlo (RMC) simulations techniques. The AWAXS measurements were performed at eight different incident 
photon energies, including some close to the Ni and Zr K edges. From the measurements eight total 
structure factor ${\cal S}(K,E)$ were derived. Using the AWAXS data four differential structure 
factors $DSF_i(K,E_m,E_n)$ were derived, two about the Ni and Zr edges. The partial structure 
factors ${\cal S}_{\text{Ni-Ni}}(K)$, ${\cal S}_{\text{Ni-Zr}}(K)$ and ${\cal S}_{\text{Zr-Zr}}(K)$ 
were estimated by using two different methods. First, the ${\cal S}(K,E)$ and $DSF_i(K,E_m,E_n)$ 
factors were combined and used in a matrix inversion process. Second, three ${\cal S}(K,E)$ factors 
were used as input data in the RMC technique. The coordination numbers and interatomic distances for 
the first neighbors extracted from the partial structure factors obtained by these two methods show 
a good agreement. By using the three-dimensional structure derived from the RMC simulations, the 
bond-angle distributions were calculated and they suggest the presence of distorted triangular-faced 
polyhedral units in the amorphous NiZr$_2$ structure. We have used the Warren chemical short-range 
order parameter to evaluate the chemical short-range order for the amorphous NiZr$_2$ 
alloy and for the NiZr$_2$ compound. The calculated values show that the chemical short-range order 
found in these two materials is similar to that found in a solid solution. 

\end{abstract}


\maketitle

\section{Introduction}

Amorphous materials have a great potential for application in technological devices, but their uses 
are limited due to several factors. One of them is the difficulty to obtain information about their 
atomic structures. Since most chemical and physical properties of these materials are defined by the 
first neighbors, knowledge of their structures becomes necessary. Due to the lack of translational 
symmetry, the determination of their atomic structures involves the combination of different 
diffraction and spectroscopic techniques as well as simulations and modeling. 

The structure of an amorphous binary alloy is described by three pair correlation functions 
$G_{ij}(r)$, which are the Fourier transformation of the three partial structure factors 
${\cal S}_{ij}(K)$. The total structure factor ${\cal S}(K,E)$, which can be derived from scattering 
measurements, is a weighted sum of these three ${\cal S}_{ij}(K)$ factors \cite{Faber}. Thus, to 
determine the three partial ${\cal S}_{ij}(K)$ at least three independent ${\cal S}(K,E)$ factors are 
needed. Usually, the following methods have been used to obtain these ${\cal S}(K,E)$ factors:

\begin{enumerate}
\item Isomorphous substitution: in this method one component of the amorphous 
alloy is substituted by a (expected) chemically similar one. For example, in amorphous 
Ni-Zr alloys the Zr atoms are replaced by Hf atoms or the Ni atoms are partially substituted 
by Co or Fe atoms \cite{tanner}. This method is not very accurate since one cannot be sure of the 
chemical similarity of the two elements involved.

\item Isotope substitution: in this method one component of the amorphous alloy is substituted by 
an isotope or by an isotopic mixture and the scattering measurements are 
carried out using neutron sources. For example, in amorphous Ni-Zr alloys the Ni atoms 
are replaced by a Ni isotope \cite{mizoguchi}. The advantages of this method are limited by the 
availability of adequate isotopes. 

\end{enumerate}

With the development of the synchrotron radiation sources, anomalous wide angle x-ray scattering 
(AWAXS) and differential anomalous scattering (DAS) techniques became available for structural study 
of multicomponent disordered materials. AWAXS utilizes an incident radiation that is tuned close to 
an atomic absorption edge so that it interacts resonantly with the electrons of that particular atom. 
The atomic scattering factor of each chemical component can therefore be varied individually and the 
chemical environment about each component in the material can be investigated. Thus, in the case of 
an amorphous binary alloy, by using only one sample three independent ${\cal S}(K,E)$ factors can be 
obtained. However, the matrix formed by the weights of the three ${\cal S}(K,E)$ factors is 
ill-conditioned, becoming difficult the determination of the three ${\cal S}_{ij}(K)$ factors.

Fuoss\cite{fuoss1,fuoss2} tried to overcome this difficulty by implementing the DAS approach, which 
was proposed by Schevick \cite{Shevchik1,Shevchik2}. The DAS approach consists in taking the 
difference between the scattering patterns measured at two incident photon energies just below the 
edge of a particular atom, so that all correlation not involving this atom subtract out since only the 
atomic scattering factor of this atom changes appreciably. Later, De Lima {\em et al}. 
\cite{Tonnerre}, following a suggestion made by Munro \cite{Munro}, combined the differential 
scattering factors $DSF_i(K,E_m,E_n)$ and  the ${\cal S}(K,E)$ factors. They observed that this 
combination reduces the conditioning number of the matrix formed by the weights of these factors, 
allowing more stable values of ${\cal S}_{ij}(K)$ to be obtained.

The reverse Monte Carlo (RMC) simulation techni\-que \cite{RMC1,RMC2,RMCA,rmcreview} has been 
successfully used for structural modeling of amorphous structures. ${\cal S}(K,E)$ factors or their 
Fourier transformations, the total reduced radial distribution functions $G(r)$, can be used as 
input data. Applications of this technique to polymeric \cite{Rosi}, crystalline \cite{mellergard} and 
amorphous~\cite{RMC5,RMC6,RMC7,RMC9,RMC11,RMC12,RMC13,kleber,rmcNDXRD,Iparraguirre,Ohkubo,Svab} 
materials are described in the literature. 

In this paper, we report the determination of the three ${\cal S}_{ij}(K)$ factors obtained for 
amorphous NiZr$_2$ alloy by two independent ways: by making a combination of AWAXS and DAS techniques 
and also by combining AWAXS and RMC simulations techniques.

\section{Theoretical background}
\label{secRMCtheory}

\subsection{Total and partial structure factors}
\label{sectheory}

According to Faber and Ziman \cite{Faber}, 
the ${\cal S}(K,E)$ factor is obtained 
from the scattered intensity per atom $I_a(K,E)$ as follows: 

\begin{eqnarray}
{\cal S}(K,E) &=& \frac{I_a(K,E)-\bigl[\langle f^2(K,E)\rangle - \langle f(K,E)\rangle^2
\bigr]}{\langle f(K,E)\rangle^2} \nonumber\,,\\
&=& \sum_{i=1}^n{\sum_{j=1}^n{W_{ij}(K,E) {\cal S}_{ij}(K) }}\,,
\label{eqstructurefactor}
\end{eqnarray}

\noindent where $K$ is the transferred momentum, $E$ is the incident photon energy, 
${\cal S}_{ij}(K)$ are the partial structure factors and $W_{ij}(K,E)$ are given by

\begin{equation}
W_{ij}(K,E) = \frac{c_i c_j f_i(K,E) f_j(K,E)}{\langle 
f(K,E)\rangle^2}\,,
\label{eqw}
\end{equation}

\noindent and

\begin{eqnarray}
\langle f^2(K,E) \rangle &=& \sum_{i}{ c_i f_i^2(K,E)}\nonumber\,,\\
\langle f(K,E) \rangle^2 &=& \Bigl[\sum_{i}{ c_i f_i(K,E)}\Bigr]^2 \nonumber\,.
\end{eqnarray}

\noindent Here, $c_i$ is the concentration and $f_i(K,E) = f_0(K) + f'(E) + if''(E)$ is the 
atomic scattering 
factor of the component of type $i$ and  $f'(E)$ and $f''(E)$ are the anomalous dispersion terms. 
The total reduced distribution function $G(r)$ is related to the ${\cal S}(K,E)$ factor through the 
Fourier transformation 

\begin{equation}
G(r) = \frac{2}{\pi} \int_0^{\infty}{K\bigl[{\cal S}(K,E)-1 \bigr] \sin (Kr)\, dK}\,,
\nonumber
\end{equation}

\noindent while the partial reduced distribution functions $G_{ij}(r)$ are related to the 
${\cal S}_{ij}(K)$ by means of the Fourier transformation 

\begin{equation}
G_{ij}(r) = \frac{2}{\pi} \int_0^{\infty}{K\bigl[{\cal S}_{ij}(K)-1 \bigr] 
\sin (Kr)\, dK}\,.
\nonumber
\end{equation}

From $G_{ij}(r)$ the partial radial distribution functions $\text{RDF}_{ij}(r)$ can be calculated by

\begin{equation}
\text{RDF}_{ij}(r) = 4\pi \rho_0 c_j r^2+ r G_{ij}(r)\,.
\nonumber
\end{equation}

\noindent Interatomic distances are obtained from the maxima of the $G_{ij}(r)$ 
functions and coordination numbers are calculated by integrating the peaks of the 
$\text{RDF}_{ij}(r)$ functions. 

\subsection{Differential structure factors and differential radial distribution functions}

By using the Faber and Ziman formalism \cite{Faber}, the $DSF_i(K,E_m,E_n)$ factor around the 
component $i$ is obtained as follows  

\begin{widetext}
\begin{eqnarray}
DSF_i (K,E_m,E_n) &=& \frac{\bigl[ I_a(K,E_m)-I_a(K,E_n)\bigr] -
\bigl[LS(K,E_m) - LS(K,E_n)\bigr]}{\langle f(K,E_m)\rangle^2 - \langle f(K,E_n)\rangle^2} 
\nonumber\,,\\
&=& \sum_j{U_{ij}(K,E_m,E_n) {\cal S}_{ij}(K)}\,,
\label{eqdsfpartial}
\end{eqnarray}
\end{widetext}

\noindent where 

\begin{equation}
LS(K,E) = \langle f^2(K,E)\rangle - \langle f(K,E)\rangle^2 \,,\nonumber
\end{equation}

\noindent is the Laue scattering term and

\begin{widetext}  
\begin{equation}
U_{ij}(K,E_m,E_n) = \frac{c_i c_j \bigl[
f_i(K,E_m) f_j(K,E_m) - f_i(K,E_n) f_j(K,E_n)\bigr]}{
\langle f(K,E_m)\rangle^2 -\langle f(K,E_n)\rangle^2 }\,.
\label{eqwdsf}
\end{equation}
\end{widetext}  

\noindent For $i \neq j$ the weight $U_{ij}(K,E_m,E_n)$ must be multiplied by 2. The differential 
distribution function $DDF_i(r)$ can be calculated by

\begin{widetext}  
\begin{equation}
DDF_i(r) = 4\pi \rho r^2+ 
\frac{2 r}{\pi} \int_0^{\infty}{K\bigl[DSF_i(K,E_m,E_n)-1 \bigr] \sin (Kr)\, dK}
\,.
\nonumber
\end{equation}
\end{widetext}  

\noindent For a binary amorphous alloy, as compared to the total radial distribution function 
$\text{RDF}(r)$, which is a sum of three $\text{RDF}_{ij}(r)$ functions, this function is the sum 
of only two $\text{RDF}_{ij}(r)$ functions and is sensitive only to the environment around the 
atoms of type $i$. In this way it is similar to EXAFS, though there are important differences due to 
the different sections of $K$-space measured \cite{Bienenstock}.

\subsection{The matrix inversion method}

In order to determine the three ${\cal S}_{ij}(K)$ factors using the eight ${\cal S}(K,E)$ and 
four $DSF_i(K,E_m,E_n)$ factors, a least-squares method was used. First of all, we define the 
function

\begin{equation}
{\cal F}(K) = \sum_{\ell=1}^M{P_\ell\biggl[{\cal SF}_\ell(K) - 
\sum_{i=1}^2{\sum_{j=1}^2{h_{ij}^\ell(K){\cal S}_{ij}(K)}}\biggr]^2}
\nonumber\,.
\end{equation}

\noindent For $1 \leqslant \ell \leqslant 8$, ${\cal SF}_\ell(K)$ indicates the ${\cal S}(K,E_\ell)$ 
factors and for $9 \leqslant \ell \leqslant 12$ it stands for the $DSF_i^\ell(K,E_m,E_n)$ factors, as 
defined by Eqs. \ref{eqstructurefactor} and \ref{eqdsfpartial}. In the same way, for 
$1 \leqslant \ell \leqslant 8$, $h_{ij}^\ell(K)$ indicates the $W_{ij}(K,E_\ell)$ weights and for 
$9 \leqslant \ell \leqslant 12$ it stands for the $U_{ij}^\ell(K,E_m,E_n)$ weights, as seen in 
Eqs. \ref{eqw} and \ref{eqwdsf}. $P_\ell$ is a weight which allows us to vary the contribution of 
any ${\cal SF}_\ell(K)$ factor. Here, $P_\ell = 1$, $\forall \ell$, and $M = 12$. By taking the derivatives of the 
${\cal F}(K)$ function with respect to the ${\cal S}_{ij}(K)$ factors and putting them equal zero a 
linear system formed by three equations can be found. These three linear equations can be written as 
a matrix equation through

\begin{equation}
\tilde{{\cal S}}=\tilde{W}\tilde{{\cal S}}_{ij}\,,\nonumber
\end{equation}

\noindent and by inverting this matrix equation the ${\cal S}_{ij}(K)$ can be found. There are several 
methods to invert a matrix, and we have adopted the Singular Value Decomposition (SVD) 
method \cite{Press}.

\subsection{The RMC method}
\label{secRMCm}

The basic idea and the algorithm of the standard RMC method are described 
elsewhere\cite{RMC1,RMC2,RMCA,rmcreview} and its application to different materials is reported in 
the literature \cite{Rosi,mellergard,RMC5,RMC6,RMC7,RMC9,RMC11,RMC12,RMC13,kleber,rmcNDXRD,Iparraguirre,Ohkubo,Svab}. 
In the RMC procedure, a three-dimensional arrangement of atoms with the same density and chemical 
composition of the alloy is placed into a cell (usually cubic) with periodic boundary conditions and 
the $G_{ij}^{\text{RMC}}(r)$ functions corresponding to it are directly calculated through

\begin{equation}
G^{\text{RMC}}_{ij}(r) = \frac{n^{\text{RMC}}_{ij}(r)}{4\pi\rho_0 r^2\Delta r}\,, \nonumber
\end{equation}

\noindent where $n^{\text{RMC}}_{ij}(r)$ is the number of atoms at a distance between $r$ and 
$r + \Delta r$ from the central atom, averaged over all atoms. By allowing the atoms to 
move (one at each time) inside the cell, the $G^{\text{RMC}}_{ij}(r)$ functions can be changed 
and, as a consequence, the ${\cal S}_{ij}(K)$ and ${\cal S}^{\text{RMC}}(K)$ factors are changed. 
Thus, the ${\cal S}^{\text{RMC}}(K,E)$ factor is 
compared to the ${\cal S}(K,E)$ factor in 
order to minimize the differences between them. The function to be minimized is 

\begin{equation}
\psi^2(E) = \frac{1}{\delta}\sum_{i=1}^m{\bigl[{\cal S}(K_i,E)-{\cal S}^{\text{RMC}}(K_i,E)\bigr]^2}\,,
\nonumber
\end{equation}

\noindent where the sum is over $m$ experimental points and $\delta$ is related to the 
experimental error in ${\cal S}(K,E)$. If the movement decreases $\psi^2$, it is always accepted. 
If it increases $\psi^2$, it is accepted with a probability given by $\exp(-\Delta 
\psi^2/2)$; otherwise it is rejected. As this process is iterated $\psi^2$ decreases until it 
reaches an equilibrium value. Thus, the atomic configuration corresponding to  
equilibrium should be consistent with the experimental total structure factor within the 
experimental error. By using the $G^{\text{RMC}}_{ij}(r)$ functions and the 
${\cal S}^{\text{RMC}}_{ij}(K) $  
factors the coordination numbers and 
interatomic distances can be calculated. In addition, the bond-angle distributions can 
also be determined. 

\section{Experimental Procedure}

\subsection{Sample and holder}

Amorphous NiZr$_2$ thin films were prepared on a NaCl cooled substrate by the sputtering triode 
technique at the Argonne National Laboratory. In order to eliminate the substrate a metallic holder 
with a central rectangular hole was glued on the thin film, and the set was immersed in water to 
dissolve the NaCl substrate. The thickness and density of the thin films were 2 $\mu$m and 
7.32 g/cm$^3$, respectively. 

\subsection{Apparatus and data collection}

The AWAXS scattering experiments were performed some years ago on the D4 beamline  at  
LURE (Orsay - France) using the DCI synchrotron radiation source. All measurements were performed 
using a two-circles diffractometer in the vertical plane described 
elsewhere\cite{Laridjani1,Laridjani2} equipped with a two crystal Si (220) or Ge (400) monochromator 
and a Si:Li energy sensitive detector which is sufficient to discriminate the large 
K$_\alpha$ resonant Raman/fluorescence signal when incident photon energy is tuned close to the Ni 
and Zr K edges but insufficient to distinguish the small elastic, K$_\beta$ and Compton signals. 
The Si (220) and Ge (400) crystal monochromators were used to collect data around the Ni and Zr edges, 
respectively. The sample was sealed, under vacuum, into a cell containing a large kapton window fixed 
around the diffractometer horizontal rotation axis to reduce the air scattering. In order to take into 
account the time-decrease of the beam, the incident intensity $I_0$ was measured by placing a NaI 
photomultiplier tube between the monochromator and the diffractometer. Absorption measurements were 
performed on the Ni and Zr K edges for energy calibration. For each incident photon energy listed 
in Table \ref{tabf}, the energy windows for K$_\alpha$, resonant Raman/fluorescence, K$_\beta$, 
Compton, and elastic signals were defined by measuring the scattered intensities at two different 
angular positions: the beginning of the windows was defined at the lowest scattering angle and their 
ends at the highest one. Scattering patterns were collected using the symmetrical reflection geometry. 
To improve the quality of the data, at least four scattering measurements were performed at the 
incident photon energies shown in Table \ref{tabf}.

\begin{table}[h]
\caption{\label{tabf} $f_{i}'$ and $f_i''$ values used here.}
\begin{ruledtabular}
\begin{tabular}{ccccc}
Energy (eV) & $f_{\text{Ni}}'$ & $f_{\text{Ni}}''$ & $f_{\text{Zr}}'$ & $f_{\text{Zr}}''$ \\
8210 & -3.895 & 0.491 & -0.347 & 2.167 \\
8330 & -7.333 & 0.658 & -0.367 & 2.167 \\
9659 & -0.964 & 3.051 & -0.586 & 1.629 \\
11867 & -0.063 & 2.158 & -0.950 & 1.126 \\
16101 & 0.254 & 1.287 & -1.985 & 0.648 \\
17398 & 0.269 & 1.121 & -2.999 & 0.563 \\
17987 & 0.269 & 1.057 & -7.371 & 0.734 \\
19200 & 0.264 & 0.940 & -1.953 & 3.315
\end{tabular}
\end{ruledtabular}
\end{table}

\subsection{Determination of the anomalous scattering factors}

In order to interpret the scattering data correctly, the real and the imaginary parts 
$f'$ and $f''$ of the atomic scattering factor were determined accurately following a procedure 
described by Dreier {\em et al}.\cite{Dreier}  X-ray absorption coefficients were measured over a 
broad energy range near the Ni and Zr K edges on the sample and $f''$ was calculated using the 
optical theorem. The absorption measurements were performed on the XAS beamline at LURE, in 
transmission mode, and two standard ionization chambers were used as the detecting system. Outside the 
region of measurement, theoretical values \cite{mcmaster} of $f''$ were used to extend the 
experimental data set over a larger energy range and $f'$ was calculated using the 
Kramers-Kronig relation. For the measurements away from the K edges, the $f'$ and $f''$ 
values were taken from a table compiled by Sasaki \cite{Sasaki}. The obtained values for the incident 
photon energies 8330 eV and 17987 eV are listed in Table \ref{tabf} together with Sasaki values. The 
atomic scattering factor away from the K edge, $f_0(K)$, of neutral Ni and Zr atoms were calculating 
according to the analytic function given by Cromer and Mann \cite{Cromer}.

\subsection{Data analysis}

Each scattering pattern was treated separately. Due to the diffractometer characteristics, the 
polarization correction was disregarded. The measured scattered intensities away from the Ni and Zr 
K edges were corrected for detector non-linearities and reabsorption effects being then put on a per 
atom scale and the Compton scattering eliminated. For those measured close to the Ni and Zr edges, 
besides the corrections above, the contribution of the K$_\beta$ fluorescence to the elastic 
scattering was eliminated. This contribution was evaluated, after corrections for reabsorption effects, 
by using the K$_\beta$/K$_\alpha$ ratio, which was measured at an incident photon energy above the 
edge, because they were completely separated. Two method were tested to put the scattered intensities 
on a per atom scale: that described by Krogh-Moe-Norman \cite{Krogh,Norman,Wagner} and the one based 
on the high-angle procedure \cite{Wagner,Gingrich}. For the elastic scattered intensity around the 
Zr edge, for which the $K$ data range is larger ($K_{\text{max}} \approx 13$ \AA$^{-1}$), both method 
gave the same result. The $K$ data range was reduced and the Krogh-Moe-Norman method was then 
applied. The normalized signal was identical to the one obtained previously. So, we have kept the 
Krogh-Moe-Norman method to put the scattered intensities on a per atom scale including those around 
the Ni edge, for which the $K$ data range is smaller ($K_{\text{max}} \approx 8$ \AA$^{-1}$). The 
good quality of the total structure factors shown in Fig. \ref{fig1} and \ref{fig2} 
corroborates the accuracy of the procedure used. The Compton scattering contribution was calculated 
according to the analytic approximation given by P\'alinkas \cite{Palinkas}.

\section{Results and Discussion}

\subsection{Total structure factors, differential structure factors and differential distribution 
functions}

\begin{figure}[h]
\includegraphics{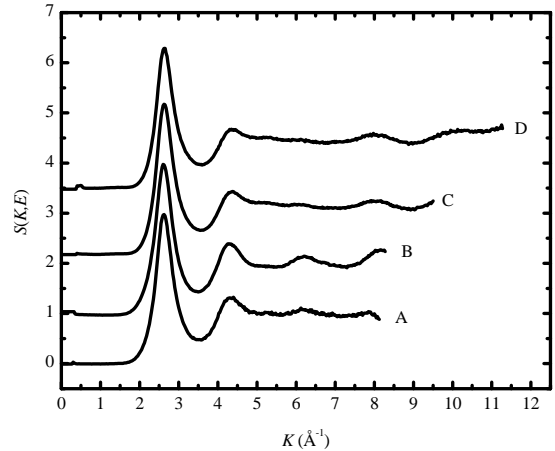}
\caption{\label{fig1} ${\cal S}(K,E)$ factors  
for energies $E_1=8210$ eV (A), $E_2=8330$ eV (B), $E_3=9659$ eV (C) and $E_4=11867$ eV (D).}
\end{figure}

\begin{figure}[h]
\includegraphics{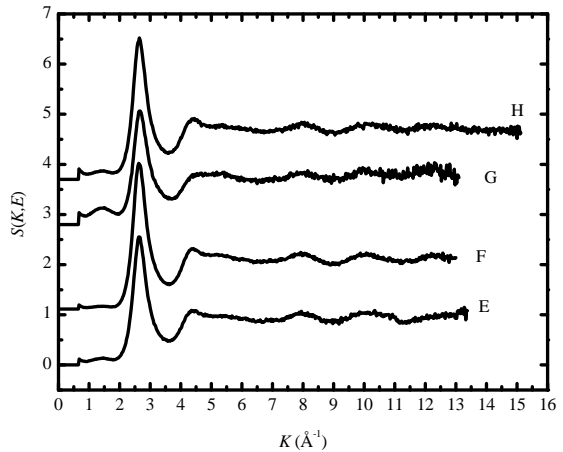}
\caption{\label{fig2} ${\cal S}(K,E)$ factors 
for energies $E_5=16101$ eV (E), $E_6=17398$ eV (F), $E_7=17987$ eV (G) and $E_8=19200$ eV (H).}
\end{figure}

Figures \ref{fig1} and \ref{fig2} show the ${\cal S}(K,E)$ factors for the incident photon energies 
described in Table \ref{tabf}. The ${\cal S}(K,E)$ factors for the energies greater than 16101 eV 
(see Fig. \ref{fig2}) show a pre-peak located at about $K = 1.65 $ \AA$^{-1}$, which reaches the 
maximum intensity close to the Zr edge. It is absent in the ${\cal S}(K,E)$ factors for the smallest 
energies (see Fig. \ref{fig1}). Since ${\cal S}(K,E)$ is a weighted sum of ${\cal S}_{ij}(K)$, we 
calculated the weights $W_{ij}(K,E)$ for the ${\cal S}_{\text{Ni-Ni}}(K)$, ${\cal S}_{\text{Ni-Zr}}(K)$ 
and ${\cal S}_{\text{Zr-Zr}}(K)$ factors. Their contributions for the ${\cal S}(K,E)$ factor at 
8330 eV are about 4\%,  32\%  and 64\% while for the ${\cal S}(K,E)$ factor at 17987 eV they are 
about 10\%, 43\% and 47\%, respectively. Thus, the determination of Ni-Ni first neighbors is more 
difficult than for the Ni-Zr and Zr-Zr first neighbors.

\begin{figure}[h]
\includegraphics{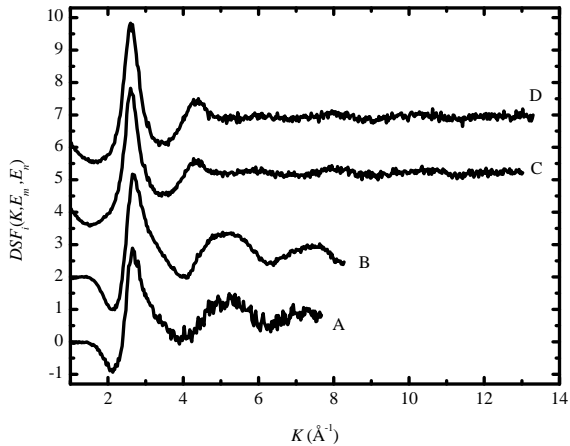}
\caption{\label{fig3} $DSF_{i}(K,E_m,E_n)$ factors. 
$DSF_{\text{Ni}}(K,E_m,E_n)$ for energies ($E_1, E_2$) (A) and ($E_2, E_3$) (B).  
$DSF_{\text{Zr}}(K,E_m,E_n)$ for energies ($E_5,E_7$) (C) and ($E_7, E_8$) (D).}
\end{figure}

\begin{figure}
\includegraphics{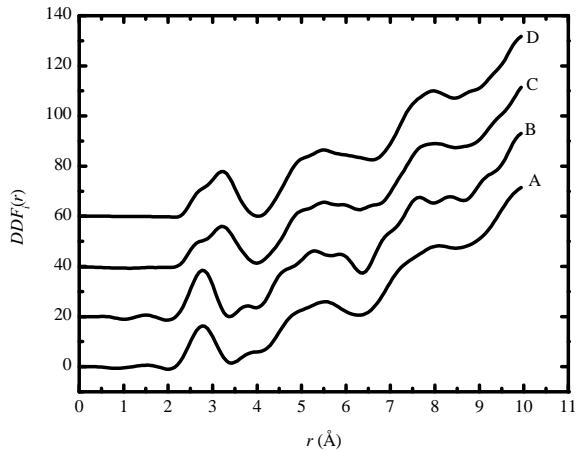}
\caption{\label{fig4} $DDF_{i}(r)$ functions. 
$DDF_{\text{Ni}}(r)$ for energies ($E_1,E_2$) (A) and ($E_2,E_3$) (B).  
$DDF_{\text{Zr}}(r)$ for energies ($E_5,E_7$) (C) and ($E_7, E_8$) (D).}
\end{figure}

We have used the scattered intensities on a per atom scale obtained at the incident photon energy 
pairs (8210, 8330 eV), (8330, 9659 eV), (16101, 17987 eV) and (17987, 19200 eV) to calculate four 
$DSF_i(K,E_m,E_n)$ factors around the Ni and Zr atoms. Figures \ref{fig3} and \ref{fig4} show them and 
their $DDF_i(r)$ functions. Figure \ref{fig3} shows that the chemical environment around Ni and Zr 
atoms is very different. For instance, it is seen a broad shoulder on the right side of the first peak 
in the $DSF_{\text{Ni}}(K,E_m,E_n)$ factors which is absent in the $DSF_{\text{Zr}}(K,E_m,E_n)$ 
factors. The $DDF_{\text{Zr}}(r)$ functions show the first coordination shell split into two 
subshells, located at about $r =  2.64$ \AA\ and 3.23 \AA, while the $DDF_{\text{Ni}}(r)$ functions 
display the first coordination shell located at a mean distance $r =  2.77$ \AA. The contribution of 
the Ni-Ni and Ni-Zr pairs to them is not resolved. This can be related to the small $K$ data range of 
the measured scattered intensities. The weight $U_{\text{Ni-Ni}}(K)$ to the $DSF_{\text{Ni}}(K,E_m,E_n)$ 
factor is about 25\%, and it is almost four times larger than the weight $W_{\text{Ni-Ni}}(K)$ for 
the ${\cal S}(K,E)$ factors, allowing a more realistic determination of the number of Ni-Ni first 
neighbors.  

By considering the atomic radius of Ni and Zr atoms, the subshells located at the distances 
$r = 2.64 $ \AA\ and 3.23 \AA\ in the $DDF_{\text{Zr}}(r)$ functions were directly attributed to 
Zr-Ni and Zr-Zr correlations. This fact allowed us to deconvolute their first shells and, 
consequently, also the first shells of the $DDF_{\text{Ni}}(r)$ functions, by assuming a Gaussian 
shape for them. To do this, we explicitly took into account the $K$ dependence by Fourier transforming 
into the $K$ space the Gaussian distributions multiplied by the weights $U_{ij}(K,E_m,E_n)$ and back 
Fourier transforming over the same $K$ range used to calculate the $DDF_i(r)$. The structural 
parameters extracted from the deconvolution are listed in Table \ref{tab2}.

\begingroup
\squeezetable
\begin{table}
\caption{\label{tab2} Structural parameters determined for the {\em a}-NiZr$_{2}$ alloy.}
\begin{ruledtabular}
\begin{tabular}{ccccccccccc}
\multicolumn{11}{c}{}\\[-0.4cm]
\multicolumn{11}{c}{Deconvolution of $DDF_i(r)$ functions} \\[0.1cm]\hline\hline
& \multicolumn{2}{c}{} & \multicolumn{2}{c}{} & \multicolumn{2}{c}{} & 
\multicolumn{4}{c}{}\\[-0.3cm]
Bond Type & \multicolumn{2}{c}{Ni-Ni} & \multicolumn{2}{c}{Ni-Zr} & \multicolumn{2}{c}{Zr-Ni} & 
\multicolumn{4}{c}{Zr-Zr}\\[0.1cm]
$N$ & \multicolumn{2}{c}{1.3} & \multicolumn{2}{c}{8.4} & \multicolumn{2}{c}{4.2} & 
\multicolumn{4}{c}{11.6} \\
$r$ (\AA) & \multicolumn{2}{c}{2.67} & \multicolumn{2}{c}{2.67} & \multicolumn{2}{c}{2.76} & 
\multicolumn{4}{c}{3.21}\\\hline\hline
\multicolumn{11}{c}{}\\[-0.3cm]
\multicolumn{11}{c}{Matrix inversion} \\[0.1cm]\hline\hline
& \multicolumn{2}{c}{} & \multicolumn{2}{c}{} & \multicolumn{2}{c}{} & 
\multicolumn{4}{c}{}\\[-0.3cm]
Bond Type & \multicolumn{2}{c}{Ni-Ni} & \multicolumn{2}{c}{Ni-Zr} & \multicolumn{2}{c}{Zr-Ni} & 
\multicolumn{4}{c}{Zr-Zr}\\[0.1cm]
$N$ & \multicolumn{2}{c}{2.3} & \multicolumn{2}{c}{7.5} & \multicolumn{2}{c}{3.8} & 
\multicolumn{4}{c}{10.8} \\
$r$ (\AA) & \multicolumn{2}{c}{2.64} & \multicolumn{2}{c}{2.77} & \multicolumn{2}{c}{2.77} & 
\multicolumn{4}{c}{3.24}\\\hline\hline
\multicolumn{11}{c}{}\\[-0.3cm]
\multicolumn{11}{c}{RMC} \\[0.1cm]\hline\hline
& \multicolumn{2}{c}{} & \multicolumn{2}{c}{} & \multicolumn{2}{c}{} & 
\multicolumn{4}{c}{}\\[-0.3cm]
Bond Type & \multicolumn{2}{c}{Ni-Ni} & 
\multicolumn{2}{c}{Ni-Zr\footnote{There are 9.2 Ni-Zr pairs at $\langle r \rangle = 2.95$ \AA.}} 
& \multicolumn{2}{c}{Zr-Ni\footnote{There are 4.6 Zr-Ni pairs at $\langle r \rangle = 2.95$ \AA.}} & 
\multicolumn{4}{c}{Zr-Zr}\\[0.1cm]
$N$ & \multicolumn{2}{c}{3.2 } & 6.9 & 2.3 & 3.5 & 1.1 & 
\multicolumn{4}{c}{10.1 } \\
$r$ (\AA) & \multicolumn{2}{c}{2.68 } & 2.73 & 3.62 & 2.73 & 3.62 
& \multicolumn{4}{c}{3.25 }\\\hline\hline
\multicolumn{11}{c}{}\\[-0.3cm]
\multicolumn{11}{c}{Crystalline NiZr$_2$ compound} \\[0.1cm]\hline\hline
& \multicolumn{2}{c}{} & \multicolumn{2}{c}{} & \multicolumn{2}{c}{} & 
\multicolumn{4}{c}{}\\[-0.3cm]
Bond Type & \multicolumn{2}{c}{Ni-Ni} & \multicolumn{2}{c}{Ni-Zr} & \multicolumn{2}{c}{Zr-Ni} & 
\multicolumn{4}{c}{Zr-Zr\footnote{There are 11 Zr-Zr pairs at $\langle r \rangle = 3.29$ \AA.}}\\[0.1cm]
$N$ & \multicolumn{2}{c}{2.0} & \multicolumn{2}{c}{8.0} & \multicolumn{2}{c}{4.0} & 
1.0 & 2.0 & 4.0 & 4.0 \\
$r$ (\AA) & \multicolumn{2}{c}{2.63} & \multicolumn{2}{c}{2.79} & \multicolumn{2}{c}{2.79} 
& 2.82 & 3.17 & 3.30 & 3.47\\\hline\hline
\multicolumn{11}{c}{}\\[-0.3cm]
\multicolumn{11}{c}{Amorphous Ni$_{35}$Zr$_{65}$ alloy \cite{tanner}} \\[0.1cm]\hline\hline
& \multicolumn{2}{c}{} & \multicolumn{2}{c}{} & \multicolumn{2}{c}{} & 
\multicolumn{4}{c}{}\\[-0.3cm]
Bond Type & \multicolumn{2}{c}{Ni-Ni} & \multicolumn{2}{c}{Ni-Zr} & \multicolumn{2}{c}{Zr-Ni} & 
\multicolumn{4}{c}{Zr-Zr}\\[0.1cm]
$N$ & \multicolumn{2}{c}{3.3} & \multicolumn{2}{c}{8.6} & \multicolumn{2}{c}{4.8} & 
\multicolumn{4}{c}{11.0} \\
$r$ (\AA) & \multicolumn{2}{c}{2.45} & \multicolumn{2}{c}{2.85} & \multicolumn{2}{c}{2.85} 
& \multicolumn{4}{c}{3.30}\\\hline\hline
\multicolumn{11}{c}{}\\[-0.3cm]
\multicolumn{11}{c}{Amorphous Ni$_{36}$Zr$_{64}$ alloy \cite{mizoguchi}} \\[0.1cm]\hline\hline
& \multicolumn{2}{c}{} & \multicolumn{2}{c}{} & \multicolumn{2}{c}{} & 
\multicolumn{4}{c}{}\\[-0.3cm]
Bond Type & \multicolumn{2}{c}{Ni-Ni} & \multicolumn{2}{c}{Ni-Zr} & \multicolumn{2}{c}{Zr-Ni} & 
\multicolumn{4}{c}{Zr-Zr}\\[0.1cm]
$N$ & \multicolumn{2}{c}{2.3} & \multicolumn{2}{c}{7.9} & \multicolumn{2}{c}{3.9} & 
\multicolumn{4}{c}{9.1} \\
$r$ (\AA) & \multicolumn{2}{c}{2.66} & \multicolumn{2}{c}{2.69} & \multicolumn{2}{c}{2.69} 
& \multicolumn{4}{c}{3.15}
\end{tabular}
\end{ruledtabular}
\end{table}
\endgroup

\subsection{Partial structure factors obtained from the matrix inversion and RMC methods}

\subsubsection{Partial ${\cal S}_{ij}(K)$ factors obtained from the matrix inversion method}

\begin{figure}[h]
\includegraphics{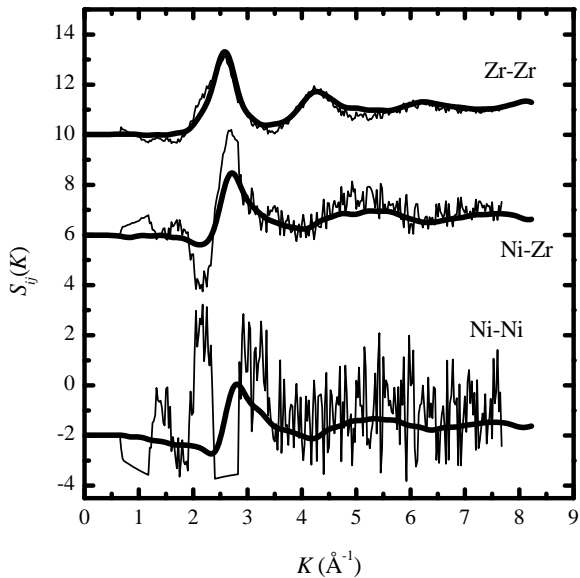}
\caption{\label{fig5} ${\cal S}_{\text{Ni-Ni}}(K)$, ${\cal S}_{\text{Ni-Zr}}(K)$ and 
${\cal S}_{\text{Zr-Zr}}(K)$ factors 
obtained from the matrix inversion (thin lines) and RMC simulations (thick lines) techniques.}
\end{figure}

By considering the eight ${\cal S}(K,E)$ and the four $DSF_i(K,E_m,E_n)$ factors, a reduction of a 
factor of 15 (at $K = 1.65$ \AA$^{-1}$) and 7 (at $K = 7.65 $ \AA$^{-1}$) in the conditioning number 
of the matrix $\tilde{W}$ was reached when compared to the use of the eight ${\cal S}(K,E)$ factors 
only. To obtain the three ${\cal S}_{ij}(K)$ factors, the ${\cal S}(K,E)$ and 
$DSF_i(K,E_m,E_n)$ factors were restricted to $K_{\text{max}}  = 7.65 $ \AA$^{-1}$.  
The ${\cal S}_{\text{Ni-Ni}}(K)$, ${\cal S}_{\text{Ni-Zr}}(K)$ and ${\cal S}_{\text{Zr-Zr}}(K)$ 
factors (thin lines) obtained are shown in Fig. \ref{fig5}. From this figure, the 
${\cal S}_{\text{Ni-Zr}}(K)$ and ${\cal S}_{\text{Zr-Zr}}(K)$ factors seem to be of good quality, 
specially between $K = 1.5$ \AA$^{-1}$ and $ K = 3.5 $ \AA$^{-1}$, where the conditioning number 
presents the largest changes. On the other hand, the ${\cal S}_{\text{Ni-Ni}}(K)$ factor displays a 
minimum apparently without physical meaning in this region. The ${\cal S}_{\text{Ni-Zr}}(K)$ factor 
shows a minimum at $K = 2.2$ \AA$^{-1}$, which was already observed in the $DSF_{\text{Ni}}(K,E_m,E_n)$ 
factors, while the ${\cal S}_{\text{Zr-Zr}}(K)$ factor displays a shoulder at this $K$ value. 
Although the ${\cal S}_{\text{Ni-Ni}}(K)$ factor is very noisy, it is possible to see a weak pre-peak 
at about $K = 1.5$ \AA$^{-1}$ which was already observed in the ${\cal S}(K,E)$ factors around the Zr 
edge and also some weak halos beyond $K = 3.5$ \AA$^{-1}$. 

\begin{figure}
\includegraphics{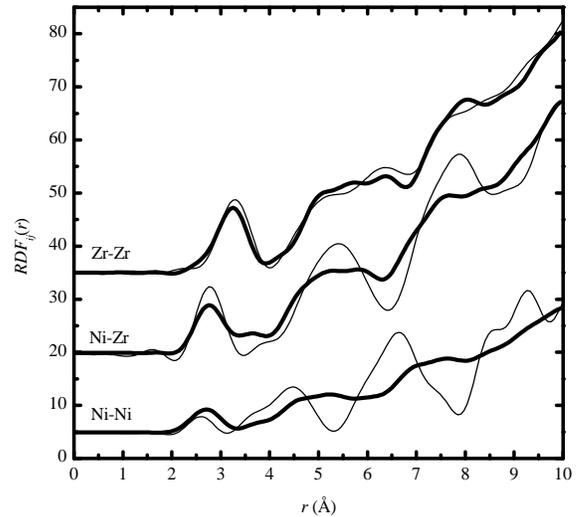}
\caption{\label{fig6} RDF$_{\text{Ni-Ni}}(r)$, RDF$_{\text{Ni-Zr}}(r)$ and RDF$_{\text{Zr-Zr}}(r)$ 
functions obtained from the matrix inversion (thin lines) and RMC simulations 
(thick lines) techniques.}
\end{figure}

Figure \ref{fig6} shows the $\text{RDF}_{ij}(r)$ functions (thin lines) obtained by Fourier 
transformation of the ${\cal S}_{ij}(K)$ factors. From this picture one can see that the 
RDF$_{\text{Ni-Ni}}(r)$ function shows a very weak first coordination shell, located at about 
$r = 2.64$ \AA, while the second and third ones, located at about $r = 4.5$ \AA\ and 6.7 \AA, 
respectively, are very intense. The RDF$_{\text{Ni-Zr}}(r)$ function displays a well isolated first 
coordination shell, located at about $r = 2.77$ \AA; a small shoulder at about $r = 3.83$ \AA, and 
well defined and intense second and third coordination shells, located at about $r = 5.4$ \AA\ and 
7.9 \AA. The RDF$_{\text{Zr-Zr}}(r)$ function shows the first coordination shell well isolated, located 
at about $r = 3.3$ \AA; the second one splits into two subshells, located at about $r = 5.18$ \AA\  
and 6.35 \AA, which are very weak. The interatomic distances and coordination numbers for the first 
neighbors obtained from these functions are listed in Table \ref{tab2}.

\subsubsection{Partial ${\cal S}_{ij}(K)$ obtained from the RMC simulations}

\begin{figure}
\includegraphics{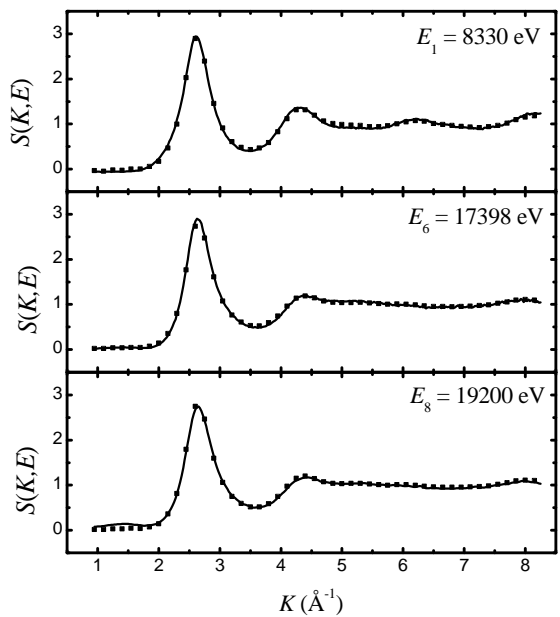}
\caption{\label{fig7} ${\cal S}(K,E)$ (full lines) and 
${\cal S}^{\text{RMC}}(K,E)$ (squares) factors for the amorphous NiZr$_{2}$ alloy.}
\end{figure}

In order to perform the simulations we have considered a cubic cell with 1800 atoms (600 Ni and 
1200 Zr), $\delta = 0.01$, and a mean atomic number density $\rho_0 = 0.054719$ atoms/\AA$^3$. The 
minimum distances between atoms were fixed at the beginning of the simulations at 
$r_{\text{Ni-Ni}} = 2.2$ \AA, $r_{\text{Ni-Zr}} = 2.4$ \AA\  and $r_{\text{Zr-Zr}} = 2.6$ \AA. 
To make the simulations we used the RMC programs available at the Internet \cite{RMCA}. 
The ${\cal S}(K,E)$ factors obtained at the photon energies 8330 eV, 17398 eV and 19200 eV were used 
as input data. These ${\cal S}(K,E)$ factors were cut at the same $K$ value. The ${\cal S}(K,E)$ 
(solid lines) and ${\cal S}^{\text{RMC}}(K,E)$ factors (square lines) are shown in Fig. \ref{fig7} 
and they show a good  agreement.

The ${\cal S}^{\text{RMC}}_{ij}(K)$ factors (thick lines) are also shown in Fig. \ref{fig5}, 
together with those obtained from the matrix inversion method. The agreement among them is very good, 
mainly for the ${\cal S}_{\text{Zr-Zr}}(K)$ factor. For instance, the shoulder previously observed 
in ${\cal S}_{\text{Zr-Zr}}(K)$ factor and the peaks located around $K = 2.58$ \AA$^{-1}$, 
4.3 \AA$^{-1}$ and 6.2 \AA$^{-1}$ are also seen in the ${\cal S}^{\text{RMC}}_{\text{Zr-Zr}}(K)$ 
factor; the minimum at $K = 2.1$ \AA$^{-1}$ and peaks around $K = 2.7$ \AA$^{-1}$,  5.2 \AA$^{-1}$ 
and 7.6 \AA$^{-1}$ observed in the ${\cal S}_{\text{Ni-Zr}}(K)$ are also seen in the the 
${\cal S}^{\text{RMC}}_{\text{Ni-Zr}}(K)$ factor; and it is interesting to note that the minimum 
previously observed  in the ${\cal S}_{\text{Ni-Ni}}(K)$ factor is now replaced by a well defined 
maximum in the ${\cal S}^{\text{RMC}}_{\text{Ni-Zr}}(K)$ factor. Other peaks observed before in the 
${\cal S}_{\text{Ni-Ni}}(K)$ factor are also seen in the ${\cal S}^{\text{RMC}}_{\text{Ni-Ni}}(K)$ 
factor. The RDF$^{\text{RMC}}_{ij}(r)$ (thick lines) functions are also shown in Fig. \ref{fig6}. 
The interatomic distances and coordination numbers for the first neighbors extracted from the 
$G^{\text{RMC}}_{ij}(r)$ and RDF$^{\text{RMC}}_{ij}(r)$ functions are also listed in Table \ref{tab2}.

It is interesting to note that although we have used eight independent ${\cal S}(K,E)$ and four 
$DSF_i(K,E_m,E_n)$ factors to obtain the three ${\cal S}_{ij}(K)$ factors by the matrix inversion 
method, the ${\cal S}_{\text{Ni-Ni}}(K)$ factor shows a minimum between $K = 1.5$ \AA$^{-1}$ and 
3.5 \AA$^{-1}$ which have no physical meaning. On the other hand, using only three ${\cal S}(K,E)$ 
factors as input data in the RMC method we have obtained excellent results, even for the 
${\cal S}_{\text{Ni-Ni}}(K)$ factor. 

In a previous study, Lee {\em et al}. \cite{tanner} and Mizoguchi {\em et al}. \cite{mizoguchi} 
reported the ${\cal S}_{ij}(K)$ factors obtained for the amorphous Ni$_{35}$Zr$_{65}$ and 
Ni$_{36}$Zr$_{64}$ alloys, respectively. Lee {\em et al} used alloys containing mixtures of Co and Fe 
in small quantities while Mizoguchi {\em et al} used the isotopic substitution method. A comparison 
among the ${\cal S}^{\text{RMC}}_{ij}(K)$, $G^{\text{RMC}}_{ij}(r)$ and RDF$^{\text{RMC}}_{ij}(r)$ 
functions obtained in this work, by using only one sample, with those reported by them shows that 
they are in a good agreement.

\begin{figure}
\includegraphics{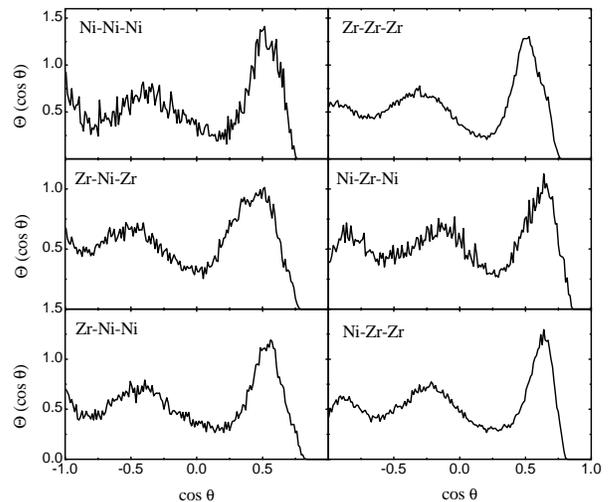}
\caption{\label{fig8} $\Theta_{i-j-l}(\cos\theta)$  functions 
obtained by the RMC simulations.}
\end{figure}

By defining the partial bond-angle distribution functions $\Theta_{i-j-l}(\cos\theta)$  
where $j$ is the atom in the corner, we calculated the angular distribution of the bonds between 
first neighbor atoms. 
The six $\Theta_{i-j-l}(\cos\theta)$ functions found are shown in Fig. \ref{fig8}. This kind of 
information cannot be obtained by the matrix inversion method because the partial distribution 
functions give only a one-dimensional description of the atomic arrangement.

Hausleitner and Hafner \cite{Hausleitner} investigated several amorphous alloys formed by transition 
metals using molecular dynamics simulations. They obtained the structure factors, coordination numbers, 
interatomic distances and also the bond-angle distribution functions. According to them, if the 
components of an alloy have a large difference in the number of $d$ electrons there is a pronounced 
non-additivity of the pair interactions and a strong interaction between unlike atoms. Therefore, 
in certain amorphous alloys the interatomic distance of heteropolar bonds can be shorter than the 
distance of homopolar ones \cite{kleber,Fukunaga}. This also favors the formation of 
trigonal-prismatic units. The shortening effect was not observed in this study. The bond angle values 
of triangular and square faces of a distorted prism are found around 
$\theta= 60^{\circ}$, $90^{\circ}$--$100^{\circ}$, $109^{\circ}$ and $147^{\circ}$. 
Thus, the bond angle values shown in Fig.~\ref{fig8} suggest that distorted polyhedral 
units with triangular faces are present in the atomic structure of the amorphous NiZr$_2$ alloy and 
if there are square-faced polyhedral units, they are found in a very small quantity. The presence 
of this kind of units was already reported in amorphous Ni$_{60}$Ti$_{40}$ \cite{kleber}, 
Ni$_{33}$Ti$_{67}$ \cite{Iparraguirre}, 
Pd$_{82}$Si$_{18}$ \cite{Ohkubo} and Ni$_{65}$B$_{35}$ \cite{Svab} alloys.

Havinga {\em et al}. \cite{Havinga} reported crystallographic data for several compounds with 
CuAl$_2$ (C16)-type structure. We have used their data for the NiZr$_2$ compound in the CRYSTAL 
OFFICE98 software\cite{Atomic} to obtain the interatomic distances and coordination numbers for the 
first neighbors listed in Table \ref{tab2}. Its structure has angles 
around $\theta = 53$--56$^\circ$, $64^\circ$, $96$--$112^\circ$ and $143$--$154^\circ$. For the 
amorphous NiZr$_2$ alloy the $\Theta_{i-\text{Zr}-l}(\cos\theta)$ functions ($i$ and 
$l$ can be Ni or Zr atoms) found are very similar to those present in the NiZr$_2$ compound. On the 
other hand, the $\Theta_{i-\text{Ni}-l}(\cos\theta)$ functions show some important differences. The 
Ni--Ni--Ni sequence in the compound is linear ($\theta = 180^\circ$) while in amorphous 
NiZr$_2$ alloy it can also be found at triangular angles ($\theta=60^\circ$ and 
$112^\circ$); the Zr--Ni--Ni and Zr--Ni--Zr sequences found in the amorphous NiZr$_2$ alloy show bond 
angles around $\theta=180^\circ$ that are not found in the compound. These data suggest that in the 
amorphous alloy the local structure around a Ni atom can be more distorted than around a Zr atom.

The chemical short-range order parameter is a quantitative measurement of the degree of chemical 
short-range order. There are several definitions of chemical short-range order 
parameters \cite{Lamparter}. We have adopted the Warren chemical short-range order (CSRO) parameter 
$\alpha_1$ \cite{Warren}  given below to compare the amorphous NiZr$_2$ alloy and the NiZr$_2$ 
compound.

\begin{equation}
\alpha_1 = 1 - \frac{N_{12}^1}{c_2(c_1N_2^1 + c_2 N_1^1)}\,,
\end{equation}

Here, $N_{12}^1 = N_{\text{Ni-Zr}}$ and 
$N_1^1 = N_{\text{Ni-Ni}} + N_{\text{Ni-Zr}}$ and $N^1_2 = N_{\text{Zr-Zr}} + N_{\text{Zr-Ni}}$ 
are the total coordination numbers in the first shell. By using the coordination numbers given in 
Table \ref{tab2}, we found $\alpha_1^{\text{RMC}} = -0.029$ and for the compound 
$\alpha_1^{\text{NiZr}_2} = -0.024$. The chemical short-range order in the amorphous NiZr$_2$ alloy 
is similar to that found in a solid solution, for which $\alpha_1 = 0$. The same behavior is 
observed in the NiZr$_2$ compound. 

\section{Conclusion}

The amorphous NiZr$_2$ alloy was investigated by using AWAXS, DAS and RMC simulations techniques. 
The three ${\cal S}_{ij}(K)$ factors were determined by using two independent ways: a combination of 
AWAXS and DAS techniques and by combining AWAXS and RMC simulations techniques. The 
${\cal S}_{\text{Ni-Zr}}(K)$ and ${\cal S}_{\text{Zr-Zr}}(K)$ factors obtained by both combinations 
showed an excellent agreement, but the simulations gave the best result for the 
${\cal S}_{\text{Ni-Ni}}(K)$ factor. The coordination numbers and interatomic distances calculated 
from the RDF$_{ij}(r)$ functions obtained by these two combinations show a good agreement among 
themselves and with those extracted from the deconvolution of the 
$DDF_{\text{Ni}}(r)$ and $DDF_{\text{Zr}}(r)$ functions. 

The results achieved in this study suggest that the combination of the RMC simulations method and 
AWAXS can be used to obtain more reliable ${\cal S}_{ij}(K)$ factors than those obtained by the matrix 
inversion method, in particular for the Ni-Ni pairs, which have the smallest weighting factor. This 
was already observed in a neutron diffraction study performed in a CuBr alloy \cite{Pusztai2}, which is also an 
ill-conditioned problem similar to ours. In addition, this combination gives us some evidence of 
the presence of distorted polyhedral units with triangular faces in the atomic structure of the 
amorphous NiZr$_2$ alloy. 

The calculated Warren CSRO parameter indicated that the chemical short-range order found in the 
amorphous NiZr$_2$ alloy is similar to that found in a solid solution, and there is some resemblance 
with that found in the NiZr$_2$ compound.

\acknowledgments

Two of the authors, J. C. De Lima and K. D. Machado, would like to thank the Conselho Nacional de 
Desenvolvimento Cient\'{\i}fico e Tecnol\'ogico (CNPq), Brazil, for financial support.


\end{document}